\documentclass{jetpl}
\usepackage{graphicx}
\twocolumn
\lat

\title{Unconventional magnetoresistance in long InSb nanowires}

\rtitle{Unconventional magnetoresistance in long InSb nanowires}

\sodtitle{Unconventional magnetoresistance in long InSb nanowires}

\author{S.\,V.\, Zaitsev-Zotov\dag\thanks{e-mail: serzz@cplire.ru}, 
Yu.\,A.\, Kumzerov\ddag, Yu.\, A.\, Firsov\ddag, and P.\, Monceau\S}

\rauthor{S.\,V.\, Zaitsev-Zotov, Yu.\,A.\, Kumzerov, Yu.\,A.\, Firsov, and
P.\, Monceau}

\sodauthor{Zaitsev-Zotov, Kumzerov, Firsov, Monceau}

\address{\dag Institute of Radioengineering and
Electronics of RAS, Mokhovaya 11,
101999 Moscow, Russia\\~\\
\ddag A.F.\,Ioffe Physical-Technical Institute of
RAS, Polytekhnicheskaya  26,
194021 Sankt-Petersburg, Russia\\~\\
\S Centre de Recherches sur Les Tr\`{e}s Basses Temp\'{e}ratures,
C.N.R.S., B.P. 166, 38042 Grenoble C\'{e}dex 9, France}
\dates{22 December 2002}

\abstract{Magnetoresistance in long correlated nanowires of degenerate
semiconductor
InSb in asbestos matrix (wire diameter of around 5 nm, length 0.1 - 1\,mm) is
studied over temperature range 2.3 - 300\,K. At zero magnetic field the electric
conduction $G$ and the current-voltage characteristics of such wires obey the
power laws $G\propto T^\alpha$, $I\propto V^\beta$, expected
for one-dimensional electron systems.  The effect of magnetic field corresponds
to a 20\% growth of the exponents $\alpha$, $\beta$ at $H=10$\,T.  The
observed magnetoresistance is caused by the magnetic-field-induced breaking of
the spin-charge separation and represents a novel mechanism of
magnetoresistance. }

\begin{document}
\maketitle
Electron-electron correlation effects being negligible in three-dimensional case
play a dominant role in one dimension.  One of the most significant consequence
of the correlation effect is the absence of quasiparticle excitations in 1D
metals.  Instead, in 1D case the collective excitations associated with separate
spin and charge degrees of freedom are developed and lead to the formation of
the so-called Luttinger liquid (LL) \cite{Rev}.  The spin-charge separation
mentioned above means different velocities for collective charge and spin
excitations.  Charge transport in LL is of collective nature and cannot be
described by the conventional kinetic equations.  A charged impurity in a 1D
electron system forms a tunneling barrier.  The absence of single-particle
excitations complicates the tunneling of electrons in LL and leads to a power
law dependence of tunneling density of states.  Tunneling through this barrier
in the case of short-range e-e interaction provides the power laws for the
linear conduction \cite{KF} $G(T)=G_0\left(T/\epsilon\right)^\alpha$ and
nonlinear I-V curve $ I(V)=I_0\left(V/V_0\right)^\beta$, whereas for long-range
Coulomb interaction a substantially different functional dependence of type
$G\propto \exp\left[-\nu\ln(T_0/T)^{1/3}\right]$ is predicted \cite{GRS,FGS}.

Experimental study of 1D behavior remains a challenge.  Single-wall and
multi-wall carbon nanotubes and various nanowires have been intensively studied
last years (see e.g.~\cite{carbon,Bi,Sb} and references therein).  One of the
most dramatical effect of reduced dimensionality on physical properties of long
nanowires was reported recently for the electric conduction of InSb nanowires in
an asbestos matrix \cite{ZZKFM}.  It was found that $G$ as a function of
temperature and electric field follows power laws over 5 orders of magnitude of
conduction variation.  The effect was considered as a manifestation of the
Luttinger-liquid-like behavior of an impure 1D electron system.  This conclusion
has been supported recently by the measurements of thermoelectric
power~\cite{ThermoEP}.  Namely, it was found that the Seebec coefficient of InSb
nanowires as a function of temperature exhibits a metallic behavior
corresponding to n-type conduction, whereas the temperature variation of the
electric conduction follows a power law.  LL is the only known physical system
where these both types of behavior coexist~\cite{LLZeebec}.

The physical reason for realization of the LL-like behavior in InSb nanowires is
a lucky coincidence of numerous factors~\cite{ZZKFM}.  Namely, a very small
effective electron mass intrinsic to bulk InSb ($m^* \sim 10^{-2}m_e$) is
favorable for a pronounced energy level splitting due to quantum size effect,
which was estimated to $10^4$ K in our nanowires having 5 nm in
diameter~\cite{ZZKFM}.  Studied samples consist of about $10^6$ of such parallel
crystalline \cite{InSbCryst} InSb nanowires forming hexagonal a lattice with a 30 nm period.  Thus,
long-range interactions between electrons in each wire may be screened through
the Coulomb interaction of these electrons with electrons on neighboring wires.
This leads to a short-range intra-wire e-e interactions, which is a basic
assumption of the LL theory.  In all other respects, the wires can be considered
independent of each other.  It was argued~\cite{ZZKFM} that the transport
properties of individual wires are determined by tunneling through impurities
and weak links (e.g.  constrictions) introduced during the fabrication process.
Thus, such an impure LL may be considered \cite{ZZKFM} as broken into drops of
almost pure LL separated by weak links.  ``Almost pure'' means that the size of
most of the drops are less than the localization length $L_{loc}$.  If one takes
into account the repulsive e-e interaction $L_{loc}$ may be larger than the mean
distance between impurities and the mean free path (weak pinning)
\cite{Furasaki,Maurey}.  As the mean drop size is $\sim 10^3$\,nm 
\cite{ZZKFM}, this condition can be fulfilled.  Then the
dominant role in transport is determined by tunneling through weak links which
are connected in series by LL drops as it was proposed in \cite{RA}.  In
addition, due to the high density of weak links along each nanowire, contact
effects play a negligible role in transport properties.

Study of the magnetoresistance is a powerful tool for investigation of the
transport mechanism in physical systems.  In the case of LL where charge and
spin degrees of freedom are decoupled, magnetic field effect may bring new
features which are not observed in other physical systems.  In particular, it
causes breaking of the spin-charge separation~\cite{Kimura}.  As a result, the
charge mode responsible for electric conduction gets a contribution from the
spin mode.  This effect is expected to be most pronounced in TDOS and affects
the critical indixes~\cite{RabelloSi}.  So a novel type of magnetoresistance can
be expected in physical systems exhibiting a LL-like behavior whose conduction
is dominating by tunneling.  In this context InSb nanowires are very perspective
objects because of their exceptionally strong spin-orbit coupling which leads to
a very high g-factor $g\approx 50$.  Below we present magnetoresistance data
for InSb nanowires in an asbestos matrix in magnetic fields up to 10 T. The
observed magnetoresistance corresponds to a magnetic-field dependence of the
exponents $\alpha, \beta$ caused by magnetic-field-induced breaking of the
spin-charge separation, representing a novel mechanism of magnetoresistance.

We have studied the electrical conduction of long InSb nanowires crystallized
inside an asbestos matrix as a function of magnetic field, temperature and
electric field.  Sample preparation and characterization have been described in
details elsewhere~\cite{ZZKFM,InSbCryst}.  The data reported below were obtained 
for two representative samples demonstrating a LL-like behavior with zero
magnetic field
exponents $\alpha=2.2$, $\beta=2.1$ (sample 1), $\alpha=4.5$, $\beta=4.3$
(sample~2).  Magnetoresistance of both samples demonstrates a similar behavior.

\begin{figure}
\includegraphics[width=8cm]{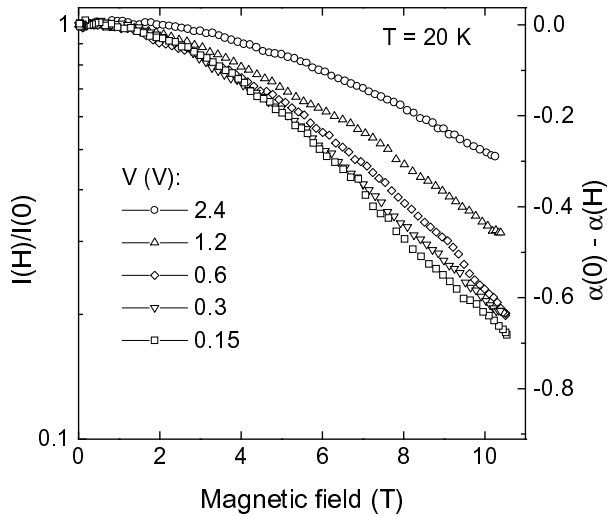}
\caption{Fig.1. Typical set of $I(H)$ curves measured at different voltages.
The right axis scale corresponds to the estimate
$\alpha(0)-\alpha(H)=ln[I(T,H)/(I(T,0)]/\ln(\epsilon/T)$ at $T=20$\,K.
 $H\perp I$. Sample 2, $\epsilon=250$\,K.}
\label{fig:IH} \end{figure}

Fig.~\ref{fig:IH} shows a typical variation of the electric current, $I$,
measured at a set of fixed voltages, {\it vs.} magnetic field.
Magnetoresistance in low magnetic fields is negative, like observed in a network
of single-wall carbon nanotubes~\cite{Nanotubes(H)}.  Conduction reaches a
maximum at $H\approx 1$\,T and falls down by a factor of 5 at $H=10$\,T at a given
temperature.  For relatively small voltages (corresponding to the linear
conduction regime) the effect of magnetic field on the conduction is practically
independent of the voltage.  For larger voltages the magnetoresistance is
smaller.  The observed magnetoconduction is slightly anisotropic and the
ratio $[I(H)-I(0)]/I(0)$ for $H\perp I$ is approximately 20\% bigger than for
$H\parallel I$.

\begin{figure}
\includegraphics[width=7cm]{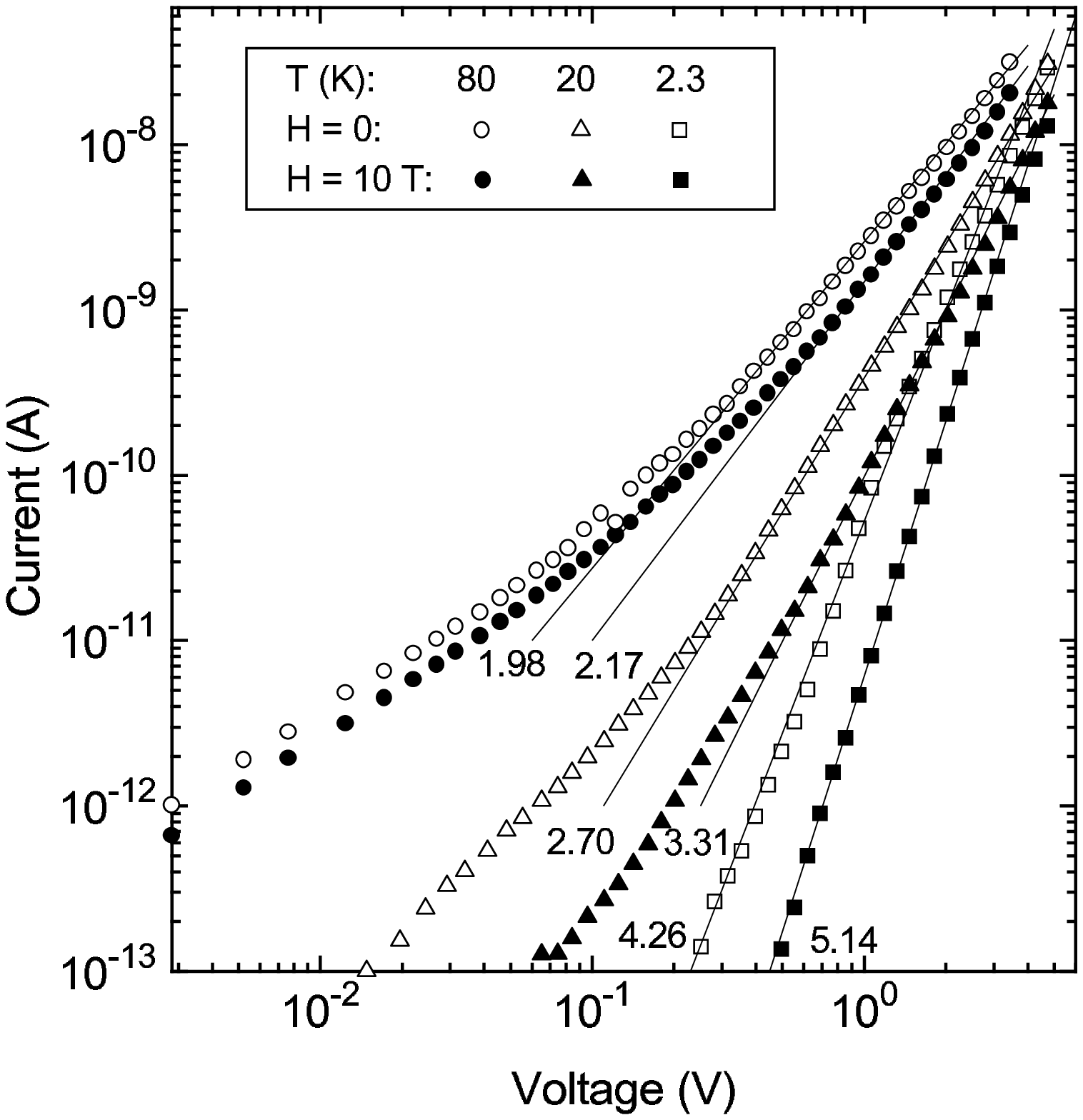}
\caption{Fig. 2. Temperature set of I-V curves measured at zero magnetic field
(empty patterns) and at $H=10$\,T (filled patterns). $H\perp I$. Sample 2. 
Solid lines show the best fit of the nonlinear part by the power law, $I\propto 
V^\beta$. The respective exponents, $\beta$, are indicated by numbers.} 
\label{fig:IVH} \end{figure}

Fig.~\ref{fig:IVH} shows the effect of the magnetic field on the shape of I-V
curves.  At relatively high temperatures (80 K curve) the effect of magnetic
field is small.  At lower temperatures the effect is much more pronounced (up to
one order of magnitude at $T\leq5$\,K for $H>10$\,T) and is smaller for bigger
electric fields.  For comparison:  measurements on InSb extracted from asbestos
cracks showed only a 20\% negative magnetoconduction $G(0)-G(H)\propto H^2$ at
$T=4.2$\,K and $H=10$\,T, and magnetoresistance of InSb in vycor glass with a 7\,nm
channel network~\cite{Berezovets} is negligibly small.  The magnetoresistance of
a network of carbon nanotubes is also small~\cite{CNetwork}.  These results
demonstrate the decisive role of the one-dimensional sample topology.  In
addition, as seen from the low-temperature curves in Fig.~\ref{fig:IVH}, the
magnetic field changes the slope of the nonlinear part of the curve, i.e.
affects the exponent $\beta$.

\begin{figure}
\includegraphics[width=7cm]{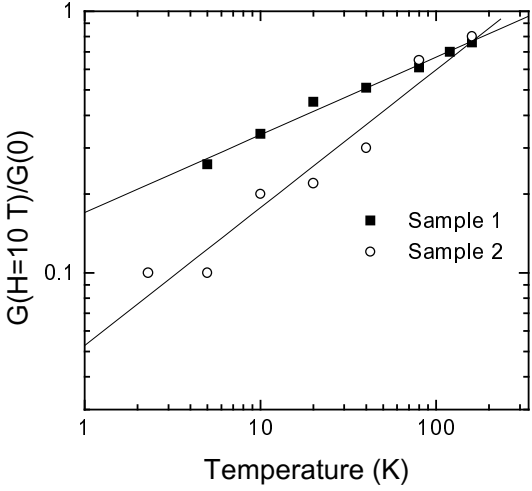}
\caption{Fig.3. Temperature dependence of the conduction in a
magnetic field of 10\,T related to its value at $H=0$
measured at the smallest $V$. Lines indicate the least-squares
fit of the data using the power law (Eq.~1). $H\perp I$.}
\label{fig:GHT}
\end{figure}

Fig.~\ref{fig:GHT} shows the temperature variation
of the ratio $G(H=10{\rm\,T})/G(0)$.  For sample 2 at $T<20$\,K this 
ratio was estimated from the low-current part of IV curves.  It is
clearly seen that the magnetoconduction depends on temperature and grows with
lowering temperature approximately as a power function of temperature.

\begin{figure}
\includegraphics[width=7cm]{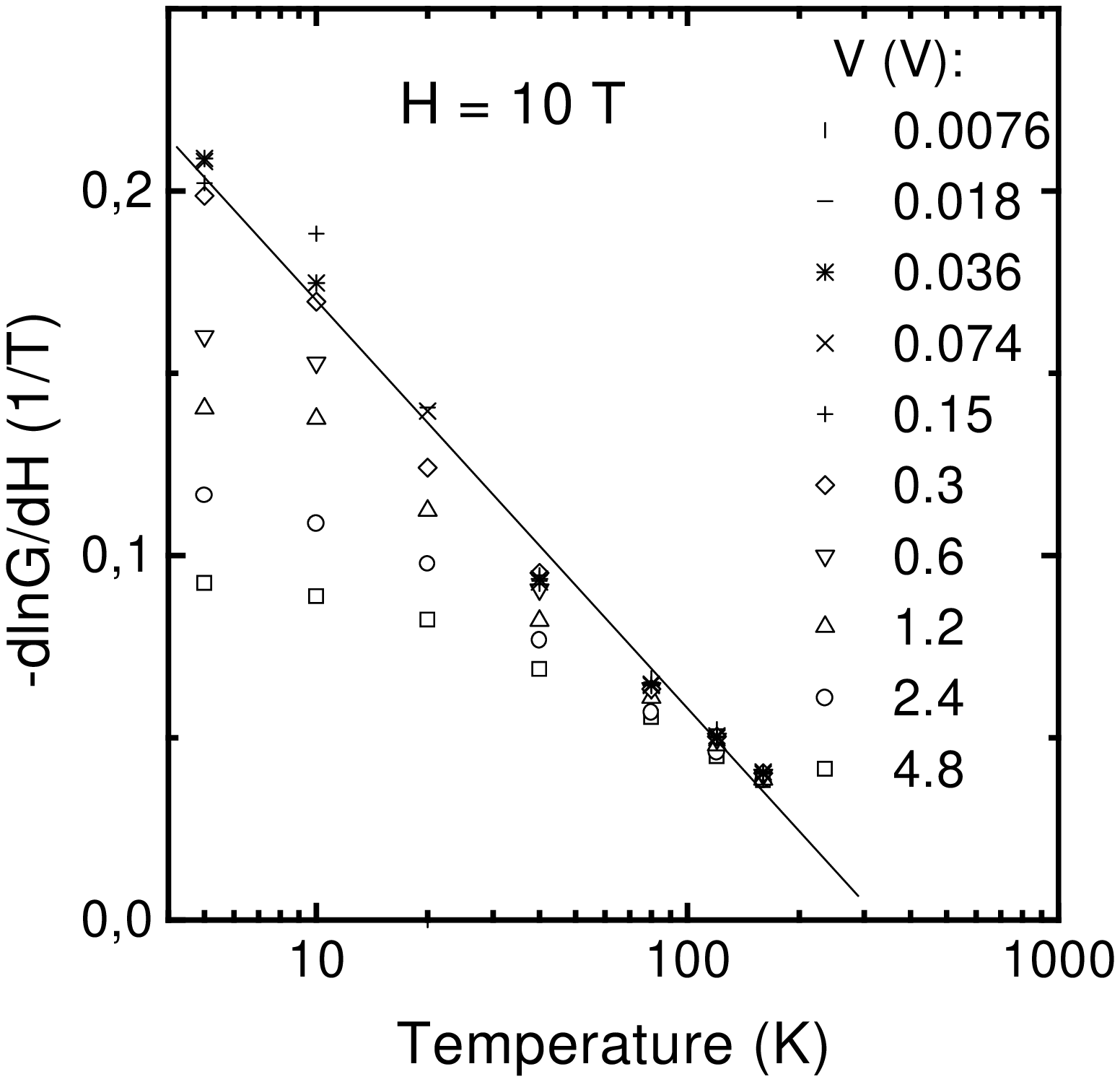}
\caption{Fig.4. Temperature dependences of the conduction
sensitivity to the magnetic field $d\ln G/dH$ at $H=10$\,T for given
values of the voltage V. The line corresponds to
the power law $d\ln G/dH=0.048 \ln(T/315{\rm\,K})$\,1/T. $H\perp I$. Sample 1.}
\label{fig:slopest}
\end{figure}

$d\ln G/dH$ at $H=10$\,T as a function of both temperature and electric field is
shown in Fig.~\ref{fig:slopest}.  At relatively small voltages $d\ln G/dH$ is
practically independent of voltage and forms a limiting dependence.  This
dependence as a function of the $\ln T$ can be approximated by a straight line,
$d\ln G/dH=A-B\ln(T)$ (solid line in Fig.~\ref{fig:slopest}).  At relatively
large voltages $d\ln G/dH$ tends to deviate from the limiting dependence upon
cooling, the deviation being growing with the voltage.

\begin{figure}
\includegraphics[height=6cm]{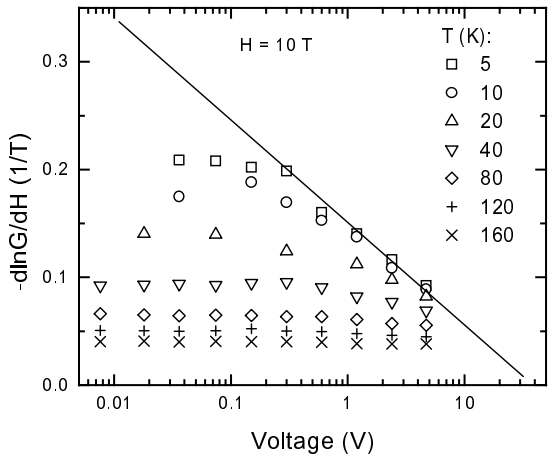}
\caption{Fig.5. Voltage dependences of the conduction sensitivity to
the magnetic field $d\ln G/dH$ at $H=10$\,T for given temperatures. The line
corresponds to
the power law $d\ln G/dH=0.041\ln(V/40{\rm\,V})$\,1/T. $H\perp I$. Sample 1.}
\label{fig:slopesv}
\end{figure}

Temperature sets of $d\ln G/dH$ plotted {\it vs.} voltage show a similar
behavior (see Fig.~\ref{fig:slopesv}).  The high-voltage data form a limiting
curve which can be approximated by $d\ln G/dH=C-D\ln(V)$ (solid line in
Fig.~\ref{fig:slopesv}).  Low-voltage data demonstrate a deviation from this
limiting curve and are practically independent of voltage at $V\rightarrow 0$.
Note that the limiting curves in Fig.~\ref{fig:slopesv} and \ref{fig:slopest}
are formed with different data:  namely, the low-voltage data forming the
limiting curve in Fig.~\ref{fig:slopest} sit on the low-voltage plateaus in
Fig.~\ref{fig:slopesv}, and vice versa, the low-temperature data forming the
limiting curve in Fig.~\ref{fig:slopesv} sit on the low-voltage plateaus in
Fig.~\ref{fig:slopest}.  Thus Figs.~\ref{fig:slopest} and \ref{fig:slopesv}
illustrate different features of the observed magnetoresistance.

As it is clear from the data of Figs.~\ref{fig:IVH}, \ref{fig:GHT},
\ref{fig:slopest}, \ref{fig:slopesv}, the observed magnetoresistance 
corresponds
to a magnetic-field induced variation of the exponents for $G(T)$
\begin{equation}
G(T)=G_0\left(\frac{T}{\epsilon}\right)^{\alpha(H)}
\label{eq:LLGTH}
\end{equation}
and nonlinear I-V curve
\begin{equation}
I(V)=I_0\left(\frac{V}{V_0}\right)^{\beta(H)}.
  \label{eq:LLIVH}
\end{equation}
Indeed, in this case $G(H)/G(0)=T^{\alpha(H)-\alpha(0)}$, in agreement with
Fig.~\ref{fig:GHT}, $d\ln G/dH=(\ln T -\ln \epsilon)d\alpha/dH$ and $d\ln
I/dH=(\ln V -\ln V_0)d\beta/dH$, in agreement with Figs.~\ref{fig:slopest} and
\ref{fig:slopesv} respectively.  Eqs.~\ref{eq:LLGTH},\ref{eq:LLIVH} fit the
data with $d\alpha /dH=0.05$\,1/T, $d\beta/dH=0.06$\,1/T and $\epsilon=335$\,K for
sample 1, and 0.11\,1/T, 0.12\,1/T and 250\,K for sample 2 at $H=10$\,T. 
$\alpha(0)-\alpha(H)$ for sample 2 calculated with Eq.~\ref{eq:LLGTH} 
and $\epsilon=250$\,K is shown in Fig.~\ref{fig:IH} (right scale).

Magnetic field affects both orbital motion and spin degree of freedom of
electrons.  While the magnetic length $L_B=\sqrt{c \hbar /eB} > d$, where $d$ is
the wire diameter, we cannot expect that orbital effects~\cite{Orbital} play a
role.  In our 5\,nm diameter nanowires this condition is broken at $B \sim 40$\,T,
far away from the maximum magnetic field 10\,T used in our measurements.  Thus we
can conclude that Zeeman splitting is responsible for the observed behavior.
This splitting in InSb is especially strong due the very big $g$-factor in InSb
($g\approx 50$).

At present there is no theory describing magnetoresistance in 1D correlated
conductors. We expect that magnetic field leads to a variation of $\alpha$
and $\beta$.  It is known, that magnetic field affects correlation function
critical exponents in the 1D Hubbard model~\cite{Frahm}, so $\alpha(H)$ and
$\beta(H)$ dependences are expected.  It is also worth to mention that
magnetic-field-dependent exponents are observed in a temperature
variation of NMR relaxation time of a spin-1/2 Heisenberg ladder
gapless phase~\cite{ladder}.  
Spin-charge separation in the LL model means that only independent collective
spin (spinons) and charge (holons) excitations are present at $H=0$.
Magnetic field acting on the spin subsystem mixes spinons and holons and
destroys thereby the spin-charge separation~\cite{Kimura}.  Magnetic-field
dependence of holon's characteristics (e.g.  velocity) caused by breaking of 
the spin-charge separation results to a variation of the exponents.  In
Ref.~\cite{RabelloSi} linear magnetic-field dependences for exponents of
spectral functions and bulk density of states, $\rho_{bulk}(\omega,H)$, are
obtained for a pure LL.  Namely, magnetic field affects the index for tunneling
density of states as 
\begin{equation}
\alpha_{bulk}(H)=\alpha_{bulk}(0)\left(1+a\frac{H}{H_c}\right),
\label{eq:Index(H)} 
\end{equation} 
where $H_c=\epsilon_F/g\mu_B$, and $a\approx
1$ in the strong coupling regime \cite{RabelloSi}.  So despite the absence of
theoretical results for $\alpha_{end}(H)$ responsible for tunneling
through a single impurity barrier in magnetic field, similar relative variation
of $\alpha, \beta$ is expected.  Then the observed 20\% variation of exponents
in a magnetic field of 10\,T corresponds to $H_c\sim 50$\,T, which is achieved at
$\epsilon_F=0.1$\,eV~\cite{ZZKFM}.

To our knowledge, a magnetoresistance of observed type and such a large 
value has not been reported yet for other physical objects exhibiting a LL-like
behavior.  Namely,
magnetoresistance of carbon nanotubes may be interpreted within the framework of
the weak localization scenario and Aharonov-Bohm effect \cite{Nanotubes(H)} and
also as due to the change in a magnetic field of the density of states at the
Fermi energy \cite{NanotubesF} for non-interacting electrons.  Reported
magnetoresistance in Bi~\cite{Bi} and Sb~\cite{Sb} nanowires of comparable
diameter ($~\sim 10$ nm) is much smaller (15\% in Bi and 0.2\% in Sb at 5~T) and
even has the opposite sign (Sb).  Strong anisotropic magnetoresistance of bulk
samples of n-InSb with electron densities $n\sim 10^{16}$\,cm$^{-3}$ is observed
in the quantum limit of applied magnetic field at $T=1.5$\,K.  In contrast with
our data, it can be explained within a conventional transport theory without
taking into account the decisive role of e-e interactions.

We would like to note that the consideration described above is valid
while $L_B > d$, i.e.  at $H<40$\,T.  When $L_B < d$, the orbital
effects~\cite{Orbital} take place.  In addition, when $H>H_c\sim 50$\,T, no spin
effect is expected any more. So the physical mechanism of magnetoresistance is
expected to be changed at $H=40$-50\,T.

Thus InSb nanowires exhibit a strong positive magnetoresistance (up to
1 order of magnitude
at $H=10$\,T) corresponding to a magnetic-field induced variation of the
exponents $\alpha$ and $\beta$ in $G\propto T^\alpha$ and $I\propto V^\beta$.
This variation is a manifistation of a novel physical mechanism of 
magnetoresistance specific for 1D systems.

S.V.Z.-Z.  is grateful to C.R.T.B.T.-C.N.R.S.  for hospitality
during the experimental part of research.  This work has been supported by
C.N.R.S.  through jumelage 19 between C.R.T.B.T.  and IRE RAS, the Russian
Foundation for Basic Research (01-02-17739, 01-02-17771, 02-02-17656), INTAS 
02-474, and by MNTP "Physics of Solid-State Nanosructures".

\end{document}